\documentstyle{article}
\vspace{1 cm}             
\def\lapproxeq{\lower .7ex\hbox{$\;\stackrel{\textstyle <}{\sim}\;$}}
\def\gapproxeq{\lower .7ex\hbox{$\;\stackrel{\textstyle >}{\sim}\;$}}

\begin{document}

\titlepage
\begin{flushright} RAL-TR-95-046 \\    
September 1995                         %
\end{flushright}                       %

\begin{center}
\vspace*{2cm}
{\large{\bf Are Diffractive Events at HERA due to a Gluemoron or a Quarkball?}}
\end{center}

\vspace*{0.75cm}
\begin{center}
F.E. Close \\
and \\
J.R. Forshaw \\
Rutherford Appleton Laboratory, \\ Chilton, Didcot OX11 0QX, England. \\
\end{center}

\vspace*{1.5cm}
\begin{abstract}
We propose a means of distinguishing between gluon- and quark-seeded
systems that may
in principle be measured in deep inelastic lepton scattering. As a particular
application, we discuss those HERA data on deep inelastic $ep$ scattering
which contain a large rapidity gap in the final state. It is standard to
interpret these
events as being due to ``pomeron" exchange, although precisely what is meant by
this remains unclear. Guided by data, we make some rather general
statements about these events and in particular we discuss the potential for
discerning whether the exchanged colour singlet system can be interpreted as
gluonic.
\end{abstract}

\newpage

In this letter we propose an experiment
that distinguishes between
glueballs and conventional quarkbased systems or ``quarkballs". The
experiment involves deep inelastic lepton scattering and we discuss
the application of these ideas  to the ``pomeron" as revealed in
the rapidity gap events which have been observed in electroproduction
at HERA \cite{rapgap}.

We shall show  on rather general grounds that in totally inclusive
leptoproduction off a hadron $H$, whose partons carry momentum fractions
$\beta$, the $Q^2$ dependence of the ``momentum" integral,
\begin{equation}
\frac{dM(Q^2)}{dQ^2} \equiv \frac{d}{dQ^2} \int_0^1 d\beta \; F_2^H(\beta,Q^2)
\end{equation}
may be used to distinguish
between gluonic or conventional (i.e. quark seeded) hadronic systems.
To the extent that the rapidity gap events at HERA enable us to
identify a pomeron structure function, i.e.
analogous to the $F_2^H(\beta,Q^2)$
in eq.(1) (and we shall discuss in some detail
the feasibility of this) we can apply these ideas to the structure of the
pomeron.

The general feature that we shall utilise is that the above
integral is related to the fraction of the target's
momentum that is carried by electrically charged constituents.
The above integral varies with
resolution: for a quarkball (such as the proton) the
integral in eq.(1) {\it falls}
gradually to an  asymptotic value whereas for a glueball it
should {\it rise} rather rapidly to its limit \cite{crr}.
The essential reason, as illustrated in the figure,
is that quarks shed momentum by
gluon bremsstrahlung whereas gluons feed momentum into $q\bar{q}$, hence in
regions of $\beta$ where quarks dominate one will have
$\frac{dF_2^H(\beta,Q^2)}{dQ^2} <0$ with
increasing $Q^2$; by contrast,  $\frac{dF_2^H(\beta,Q^2)}{dQ^2} >0$ in regions
of gluon dominance. Gluon dominance is anticipated as $\beta
\rightarrow 0$ for all systems and hence
$\frac{dF_2^H(\beta \rightarrow 0)}{dQ^2} >0$ in general. For a
quarkball this is counterbalanced by the behaviour
at large $\beta$ where valence quarks dominate:
$\frac{dF_2^H(\beta>0.3)}{dQ^2} <0$
such that in the integral $\frac{dM(Q^2)}{dQ^2} <0$ overall.
For a glueball, however, we would expect that the ``valence" gluons cause
$\frac{dF_2^H(\beta>0.3)}{dQ^2} >0$ which is quite opposite to that of a
quarkball and that a rise should be observed at essentially all
values of $\beta$ such that $\frac{dM(Q^2)}{dQ^2} >0.$

Data \cite{H1,ZEUS} on what has been widely interpreted as the pomeron
structure function \cite{IS} have been analysed in
many models. These models tend to assume either
explicitly or implicitly that the pomeron is gluonic and
propose tests based on the $\beta$-distributions of its structure function.
However, model dependent  $\beta$-distributions alone cannot confirm the
gluonic hypothesis and our $Q^2$ test is a complementary and, we suggest,
less model dependent measure.

These general remarks seem not to have been articulated in the literature
in this context; we shall now discuss the circumstances under which
they follow from the emerging HERA data and then suggest that the data already
imply a gluonic pomeron. We first
discuss what are the general implications
of the rapidity gap events which have been observed in electroproduction
at HERA \cite{rapgap} and argue that the data imply that the proton
effectively offers up a colour singlet system with which the current
interacts. We shall henceforth use
``pomeron" as shorthand for ``colour singlet that is exchanged" without
a priori prejudice as to its internal structure.

On purely kinematical grounds, events in which the incoming proton remains
intact and loses only a very small fraction of its initial momentum will be
associated with a rapidity gap. The HERA experimentalists are able to measure
this (dominant) subset of their rapidity gap events. However, it could in
principle be that the proton dissociates
into some low mass colour singlet state, rather than remaining intact.
 These events will have essentially
the same topology as the ones with a final state proton and good
instrumentation in the very forward direction is
needed in order to unravel the various components. The HERA experiments
are making good progress in this area. However, it should be realised that
data published hitherto have been corrected for the background from proton
dissociation rather than compiled by direct observation of a final state
proton.

Since any exchange of colour between the proton and photon would lead to
the production of final state hadrons which would fill in any rapidity gap
it follows that the rapidity gap events select those processes in which no
net colour is exchanged (between the photon
and proton). This is a rather general property, in
that nothing is said about the mechanism of the colour
neutralisation. For example, the recent work of Buchm\"uller and Hebecker has
the essential dynamics driven by single gluon exchange, the colour
neutralisation is supposed to occur sometime after the hard gluon interaction
but sometime before the parton hadronisation \cite{Buch}.

There has been much speculation regarding the nature of the dynamics of
diffractive DIS. These include the Regge inspired models of
refs.\cite{DL,models}, the QCD approaches of refs.\cite{QCD,NikZak,LW} and the
more novel approaches of ref.\cite{Buch,Kramer}.
Indeed, a substantial fraction of deep inelastic events were {\it predicted}
to be diffractive (i.e. contain the rapidity gap) using the simple Regge pole
ansatz for the pomeron \cite{DL}.

The cross section,
\begin{equation}
\frac{d^4 \sigma(x,Q^2,x_P,t)}{dx dQ^2 dx_P dt} = \frac{4 \pi \alpha^2}{x Q^4}
(1 - y + y^2/2) F_2^D(x,Q^2,x_P,t),
\end{equation}
defines the (dimensionful) diffractive structure function,
$F_2^D(x,Q^2,x_P,t)$, in terms of the
Bjorken-$x$ ($x$), photon virtuality ($Q^2$), momentum
transfer ($t$) and the variable $x_P$, where (ignoring the proton mass)
\begin{eqnarray}
x &=& \frac{Q^2}{Q^2 + W^2}, \nonumber \\
x_P &=& \frac{Q^2 + M_X^2 - t}{Q^2+W^2}, \nonumber \\
y &=& \frac{Q^2}{x s}.
\end{eqnarray}
The $\gamma^* p$ and $ep$ invariant masses are $W$ and $\surd s$
respectively, and $M_X$ is the invariant mass of the (observed) hadronic
system associated with the photon dissociation. The variable $x_P$ can be
interpreted as the fraction of the incident proton momentum which is carried
into the interaction.

The experiments do not yet measure $t$, rather they integrate over it and so
it is more appropriate to work with the dimensionless function
$\tilde{F}_2^D(x,Q^2,\tilde{x}_P)$ defined thus:
\begin{equation}
\frac{d^3 \sigma(x,Q^2,x_P)}{dx dQ^2 d\tilde{x}_P} =
\frac{4 \pi \alpha^2}{x Q^4}
(1 - y + y^2/2) \tilde{F}_2^D(x,Q^2,\tilde{x}_P).
\end{equation}
Note that for $-t \ll Q^2+M_X^2$, $\tilde{x}_P \approx x_P$. Since this is
usually the case, we will subsequently drop the tilde.

There are two striking empirical
properties of the observed $\tilde{F}_2^D(x,Q^2,x_P)$.
Firstly, it has only weak $Q^2$ dependence and secondly, the $x_P$
dependence can be partitioned as follows
\begin{equation}
\tilde{F}_2^D(x,Q^2,x_P) \approx h(\beta,Q^2) f(x_P),
\end{equation}
where
\begin{equation}
\beta = \frac{x}{x_P}
\end{equation}
and empirically
\begin{equation}
f(x_P) \sim x_P^{-n}
\end{equation}
where $n = 1.19 \pm 0.06 \pm 0.07$ (H1) \cite{H1} or
$n = 1.30 \pm 0.08 \stackrel{+ 0.08}{\scriptstyle{- 0.14}}$ (ZEUS) \cite{ZEUS}.
The first error is statistical and the second systematic. It should be
appreciated that the errors are large enough to admit an appreciable
non-partitioned contribution. The partitioning implies that
the production mechanism of the colour singlet is independent of the nature of
the deep inelastic probe (it remains to be seen whether the partitioning
also applies elsewhere, e.g. in diffractive photoproduction of dijets).
To the extent that a fast forward colour singlet proton or excited cluster has
been emitted, and at the level of accuracy at which partitioning
has been established, we can therefore think of the incoming proton
offering up a colour singlet which is then probed by the virtual photon.
Thus, the function $h(\beta,Q^2)$ can legitimately be referred to as the
structure function of the pomeron \cite{IS}
($\beta$ is then the Bjorken-$x$ defined in the
pomeron-photon system) and $f(x_P)$ as the pomeron flux factor.
Note that the absolute normalisations of $h(\beta,Q^2)$ and $f(x_P)$ are not
well defined, since only their product is measured.

We shall first establish with what restrictions we can impose a parton model
interpretation of the structure function, $h(\beta,Q^2)$, i.e.
\begin{equation}
h(\beta,Q^2) = \sum_{i=1}^{n_f} e_i^2 \beta \;[ q_i(\beta,Q^2) +
\bar{q}_i(\beta,Q^2)].
\end{equation}
and then draw general conclusions about its implications for the
microstructure of the colour singlet system.

The weak $Q^2$ dependence {\it implies} that
the diffractive cross section is measuring a leading twist phenomenon
and {\it suggests} that we can write the structure function
$h(\beta,Q^2)$ as a charge weighted sum over
parton (quark) densities (i.e. the
total cross section is an incoherent sum over individual elastic
$\gamma^*$-parton cross sections). However, it is important to be clear about
what is meant by these parton distribution functions.
Universality of the leading-twist parton densities
has been proven to all orders in perturbative QCD, (e.g. the quark
densities appearing in DIS are the same as those appearing in Drell-Yan)
but a key point to appreciate is that both initial and final state
interactions generally spoil the property of factorisation of
collinear divergences (and hence the concept of a parton density)
for any process that is not fully inclusive. Thus factorisation is lost as
soon as we look at anything other than the direct products of a hard
scattering; in particular {\it the measured proton in diffractive DIS will
generally spoil the factorisation properties} and hence the notion of
pomeron parton densities. However, Nikolaev and Zakharov \cite{NikZak} and
Levin and W\"usthoff \cite{LW} have demonstrated, to leading $\ln Q^2$
accuracy, that it is possible to partition the structure function as in
eq.(5) and to write $h(\beta,Q^2)$ in terms of quark probabilities as in
eq.(8). Thus, at this level of approximation, we can indeed talk about
diffractive parton densities. The possibility or otherwise of
consistently implementing next-to-leading log $Q^2$ evolution in diffractive
DIS is in our opinion an open question, due to the appearance of
factorisation breaking terms.

However, these parton densities will not be universal, since there
are interactions which are present in hadronic hard diffraction
which do not occur in diffractive DIS. The presence of an extra contribution
to hard diffraction in hadron-hadron reactions was recognised by Collins,
Frankfurt and Strikman \cite{CFS} who termed it the ``coherent pomeron''
contribution (for a more detailed study of the coherent pomeron in hard
diffraction see ref.\cite{BS}). The presence of the coherent pomeron
contribution has been suggested by the UA8 data \cite{UA8}. Further
confirmation of such a contribution (e.g. in diffractive photoproduction of
dijets) and the lack of universality of diffractive parton densities
needs to be established.

To the extent that the structure function
$h(\beta,Q^2)$ has been extracted, our test for gluonic or quark-seeded
systems may be applied to the
pomeron.

A simplistic distinction between these two broad classes of states is
that there exists some scale $\tilde{Q}_0^2$ at which all of the
momentum is carried by either quarks or gluons.
The concept of constituent parton momentum is not well defined for the
pomeron but in general we may define that for a gluonic pomeron
the second moment of the parton distributions
\begin{equation}
M_S(Q^2) = \int_0^1 d\beta \, \beta \sum_{i=1}^{n_f} (q_i(\beta,Q^2) +
\bar{q}_i(\beta,Q^2)),
\end{equation}
tends to vanish at some scale $\tilde{Q}_0^2$: $ M_S(\tilde{Q}_0^2) \rightarrow
0$.
Conversely for an intrinsically pure quark system, the second moment
of the gluon density is nugatory (at some scale). With this definition, we can
talk about constituent valence partons within the pomeron. It might be that
$\tilde{Q}_0^2$ is small such that we are in a non-perturbative regime,
or it might be that $\tilde{Q}_0^2 \gg \Lambda_{QCD}^2$, in which case the
pomeron is essentially perturbative: the distinction is not essential
for our purposes.

Ideally, we would like to measure
the total ``momentum'' carried by the quarks and anti-quarks as a function of
$Q^2$, i.e. $M_S(Q^2)$. If the pomeron is gluonic,
then the total ``momentum'' carried by the
quarks will increase with increasing $Q^2$, saturating at asymptotically high
$Q^2$ to some constant value (whose normalisation depends upon the
normalisation of $h(\beta,Q^2)$). Fortunately, this property is
preserved when we consider $M(Q^2)$, i.e. (see eq.(1))
$$ M(Q^2) = \int_0^1 d\beta \; h(\beta,Q^2). $$
The structure function can be written
as a linear combination of the singlet and non-singlet structure functions,

\begin{equation}
M(Q^2) = e_S^2 M_S(Q^2) + e_{NS}^2 M_{NS}(Q^2),
\end{equation}
where $M_{NS}(Q^2)$ can be expressed in terms of the non-singlet distribution
functions:
\begin{equation}
M_{NS}^{ij}(Q^2) = \int_0^1 d\beta\, \beta \;[q_i(\beta,Q^2) +
\bar{q}_i(\beta,Q^2) - q_j(\beta,Q^2) - \bar{q}_j(\beta,Q^2)]
\end{equation}
and $e_j^2$ is the squared charge of the $j$th quark type.
The explicit form for the charge sums and densities depends upon
the number of flavours, e.g. for $n_f=4$
\begin{equation}
M(Q^2) = \frac{1}{2}(e_u^2 +  e_d^2) M_S(Q^2) + \frac{1}{2}
(e_u^2 - e_d^2)(M_{NS}^{ud}(Q^2) + M_{NS}^{cs}(Q^2)).
\end{equation}

We recall the solutions for the second moments:
\begin{eqnarray}
M_{NS}(Q^2) &=& M_{NS}(Q_0^2) X_{NS} \nonumber \\
M_S(Q^2)    &=& M_S(Q_0^2) X_S + \frac{3 n_f}{16 + 3 n_f} (1-X_S) {\cal{N}}
\end{eqnarray}
where
\begin{eqnarray}
X_{NS} &=& {\rm exp}\left( \frac{32}{3(33-2n_f)} \ln
\frac{\alpha_s(Q^2)}{\alpha_s(Q_0^2)} \right) \nonumber \\
X_S &=& {\rm exp}\left( \frac{2(16+3n_f)}{3(33-2n_f)} \ln
\frac{\alpha_s(Q^2)}{\alpha_s(Q_0^2)} \right)
\end{eqnarray}
and
\begin{equation}
M_S(Q^2) + M_{NS}(Q^2) = {\cal{N}}.
\end{equation}
We treat ${\cal{N}}$ as an arbitrary factor, which fixes the normalisation
of $h(\beta,Q^2)$. Thus,
\begin{eqnarray}
\frac{d M(Q^2)}{d \ln Q^2} &=& \frac{X_S}{\ln Q^2/\Lambda^2}
\frac{2(16+3n_f)}{3(33-2n_f)} \left( \frac{3 {\cal{N}} n_f}{16 + 3n_f} -
e_S^2 M_S(Q_0^2) \right) \nonumber \\
&-& \frac{X_{NS}}{\ln Q^2/\Lambda^2} \frac{32}{3(33-2n_f)} e_{NS}^2
M_{NS}(Q_0^2).
\end{eqnarray}

We can now show that if
\begin{equation}
\frac{d M(Q^2)}{d\ln Q^2} > 0
\end{equation}
then the pomeron is gluonic. The truth of the above inequality implies that
\begin{equation}
M_S(Q_0^2) < \frac{3 {\cal{N}} n_f}{16 + 3n_f} \frac{1}{e_S^2} -
\frac{16}{16 + 3 n_f} \frac{X_{NS}}{X_S} \frac{e_{NS}^2}{e_S^2} M_{NS}(Q_0^2).
\end{equation}
Since $X_{NS}(Q_1^2)/X_S(Q_1^2) < X_{NS}(Q_2^2)/X_S(Q_2^2)$ for $Q_1^2 <
Q_2^2$ it therefore follows that if eq.(17) holds true at some value of $Q^2$
it must also hold at all lower values of $Q^2$. Hence there exists some
scale, $\tilde{Q}_0^2$, such that $M(\tilde{Q}_0^2) \rightarrow 0$ and
so the system is gluonic.

Since the inequality (eq.(17)) completely defines the gluonic system,
it follows that a quark-seeded state satisfies the opposite inequality, i.e.
\begin{equation}
\frac{d M(Q^2)}{d \ln Q^2} < 0.
\end{equation}
Note that our definition of quark-/gluon-seeded only requires a measurement
at some perturbative scale, $Q^2 \gg \Lambda_{QCD}^2$. It is not expected
that the detailed and essentially unknown non-perturbative physics will spoil
our conclusions.

We have implicitly assumed that the pomeron does not have a pointlike, direct
coupling to individual quarks, e.g. like the photon. Any such coupling would
automatically lead to a contribution which satisfies eq.(17) and hence could
fake a gluonic pomeron. Note that a large pointlike coupling of the pomeron
can arise if it is assumed that the pomeron interacts essentially as a single
gluon, e.g. at some scale the pomeron has a gluon density which is a
delta function at $x_P$ whilst the quark density is zero. Such a scenario
is the extreme case of a super-hard gluonic pomeron and as such will
satisfy eq.(17). This is the basic idea of the model of ref.\cite{Buch}.
Any other (i.e. anomalous) pointlike coupling would be of remarkable interest
in itself but remains at the level of speculation \cite{Kramer}.
Note that the model of Donnachie-Landshoff does allow the pomeron to have a
non-pointlike, direct coupling to quarks, i.e. coupling to off-shell quarks is
suppressed, by a form factor, at a rate which is characteristic of the
pomeron radius. Such a coupling kills off the leading logarithmic
contribution and does not, at leading log $Q^2$, affect the $Q^2$ evolution.

We recall that, to the extent that colour singlet exchange drives the
diffractive events, one may anticipate photoproduction of glueball type
hadrons in the central region. In $pp \rightarrow p_f(X)p_s$ (where $p_{f,s}$
denote forward and spectator protons) the centrally
produced $X$ is hypothesised
to be glueball rich if it is produced by pairs of colour singlet gluonic
pomerons, or ``gluemorons" \cite{WA91}.
Similar phenomena may be expected in the diffractive events at HERA if these
are due to gluemorons. Such isolated central production is less clear in
models of the Buchmuller type \cite{Buch} and the implications of this as a
possible discriminator merit further investigation.

In the case of quark hadrons, such as the proton, the sign of
$\frac{d M(Q^2)}{d \ln Q^2}$ tends to be hidden in the errors
due to the counterbalancing rise at small $x$ and fall at large $x$ where
the valence quarks dominate (fig.1(a)). A similar situation holds for the pion
and we note that it will be an interesting test of our general procedure to
isolate the pion cloud in $\gamma^* p \rightarrow n + X$.
For a gluonic system one may therefore anticipate a sharper
signal since the valence gluons will drive
a rise in $h(\beta,Q^2)$ across the {\it entire range} of $\beta$ and hence
$\frac{d M(Q^2)}{d \ln Q^2} > 0$ (fig.1(b)). The diffractive events at
HERA seem already to manifest this phenomenon (i.e. a rise across the entire
$\beta$-range) \cite{H1,ZEUS}. Although the error bars are as
yet too large to establish that the pomeron structure function is exhibiting
a rise with increasing $Q^2$ regardless of $\beta$, the data are certainly not
falling with $Q^2$ in the ``valence" region, $\beta \sim 0.65$.
 This already contrasts
with nucleon structure functions where the valence quarks are shedding
momentum even for $x$ ($``\beta"$) as small as $0.3$. Thus barring some
radically new behaviour at large $\beta$ (which would be at least as
interesting as all that has gone before) the result
$\frac{d M(Q^2)}{d \ln Q^2} > 0$ seems possibly to be established
qualitatively already. As such this could well be the first evidence for the
presence of colour singlet gluonic systems.

\end{document}